\documentclass[osajnl,twocolumn,showpacs,ocis,superscriptaddress,10pt]{revtex4-1} 

\usepackage{color}

\definecolor{darkblue}{rgb}{0,0,0.5}
\definecolor{lila}{rgb}{0.3,0,0.3}
\definecolor{turq}{rgb}{0,0.1,0.4}
\definecolor{lightblue}{rgb}{0.7,0.7,0.9}
\usepackage{url} 

\usepackage[pdftex,
 colorlinks=true,
 backref=page,
 linkcolor=darkblue, 
 filecolor=red,
 citecolor=turq, 
 urlcolor=lila, 
 pdftitle={Faraday Filtering on the Cs-D1-Line for Quantum Hybrid Systems},
 pdfauthor={Matthias Widmann and Simone Portalupi and Sang-Yun Lee and Peter Michler and Jorg Wrachtrup and Ilja Gerhardt},
 pdfsubject={},
 pdfkeywords={Faraday Filters, Quantum Dots, Quantum Hybrid Systems, Cesium, Atomic Spectroscopy},
 pdfpagelabels=true, 
 breaklinks=false,
 plainpages=false,
 backref=false,
 bookmarks,
 bookmarksnumbered=true]{hyperref}

\newcommand{\degree}{\ensuremath{^\circ}}

\usepackage{amsmath,amssymb,graphicx}

\begin{document}

\title{Faraday Filtering on the Cs-D$_1$-Line for Quantum Hybrid Systems}

\author{Matthias Widmann}
\affiliation{3. Institute of Physics, University of Stuttgart and Stuttgart Research Center of Photonic Engineering (SCoPE) and IQST, Pfaffenwaldring 57, D-70569 Stuttgart, Germany}

\author{Simone Portalupi}
\affiliation{Institut f\"ur Halbleiteroptik und Funktionelle Grenzfl\"achen (IHFG), University of Stuttgart, Allmandring 3, D-70569 Stuttgart, Germany}

\author{Sang-Yun Lee}
\affiliation{3. Institute of Physics, University of Stuttgart and Stuttgart Research Center of Photonic Engineering (SCoPE) and IQST, Pfaffenwaldring 57, D-70569 Stuttgart, Germany}

\author{Peter Michler}
\affiliation{Institut f\"ur Halbleiteroptik und Funktionelle Grenzfl\"achen (IHFG), University of Stuttgart, Allmandring 3, D-70569 Stuttgart, Germany}

\author{J\"org Wrachtrup}
\affiliation{3. Institute of Physics, University of Stuttgart and Stuttgart Research Center of Photonic Engineering (SCoPE) and IQST, Pfaffenwaldring 57, D-70569 Stuttgart, Germany}
\affiliation{Max Planck Institute for Solid State Research, Heisenbergstra\ss e 1, D-70569 Stuttgart, Germany}

\author{Ilja Gerhardt}
\email{Corresponding author: i.gerhardt@fkf.mpg.de}
\affiliation{3. Institute of Physics, University of Stuttgart and Stuttgart Research Center of Photonic Engineering (SCoPE) and IQST, Pfaffenwaldring 57, D-70569 Stuttgart, Germany}
\affiliation{Max Planck Institute for Solid State Research, Heisenbergstra\ss e 1, D-70569 Stuttgart, Germany}

\begin{abstract}Narrow-band filtering of light is widely used in optical spectroscopy. Atomic filters, which rely on the Faraday effect, allow for GHz-wide transmission spectra, which are intrinsically matched to an atomic transition. We present an experimental realization and a theoretical study of a Faraday anomalous dispersion optical filter (FADOF) based on cesium and its D$_1$-line-transition ($6^2S_{1/2}\rightarrow6^2P_{1/2}$) around 894~nm. We also present the prospects and visions for combining this filter with the single photon emission of a single quantum dot~(QD), which matches with the atomic transition.
\end{abstract}

\pacs{(230.3810) Magneto-optic systems; (250.5590) Quantum-well, -wire and -dot devices; (350.2450) Filters, absorption.; }
\maketitle 

In the last 50 years narrow-band optical filters based on the anomalous dispersion of an atomic vapor (FADOFs) were reported, and have their main application in convenient filtering of near-atom-resonant light~\cite{ohman_soa_1956,kessler_josa_1965}. Due to their high vapor pressure, all alkali atoms except lithium were investigated on their optical properties in such filters. In the early 1990s, many different atoms were studied and narrow-band filtering was described experimentally and theoretically~\cite{dick_ol_1991,menders_ol_1991}. Today, more reports extend this work~\cite{harrell_josab_2009} and find astonishing properties, such as unpredicted narrow line-widths~\cite{wang_ol_2012,zielinska_ol_2012} or treat the problem as an optimization challenge~\cite{kiefer_s_2014}.

Parallel to the studies on atomic filters, the spectroscopy of single solid-state emitters has been an active field of research. Not only that these emitters allow for nanoscopic sensing~\cite{neumann_nnl_2013}, they also enable high-rate single photon emission~\cite{michler_s_2000}. Recently, some solid-state emitters can be optically interlinked with atomic systems~\cite{akopian_np_2011,siyushev_n_2014,ulrich_prb_2014}. One goal is the hybridization among various systems, such that the optimal properties of different systems can be combined: For example, the coherence of an atomic system often outperforms the coherence of a solid state system. On the other hand single photon emission rates from a solid-state system are superior to the flux of photons from an atom or ion in a trap. A first challenge is the optical filtering of a single photon emitter (e.g.\ a quantum dot). This allows to spectrally select a single emitter, which is subsequently matched to an atomic transition.

Here we present our experimental and theoretical results on Faraday filtering with hot atomic cesium vapor. The transmission spectrum on the cesium D$_1$-line has not been reported so far. This spectrum is of major interest for the combination between quantum dots~(QDs) and atomic cesium~\cite{ulrich_prb_2014}, since high quality QDs can be routinely produced for this wavelength (894~nm). We implement a Cs-FADOF with a solenoid configuration, and with a fixed permanent magnet. The permanent magnet simplifies, and stabilizes the experimental configuration. Our report describes in detail two key experiments, which are suggested for the combination between a single QD and a Cs-FADOF. 

\begin{figure}[hb]
  \includegraphics[width=0.9\columnwidth]{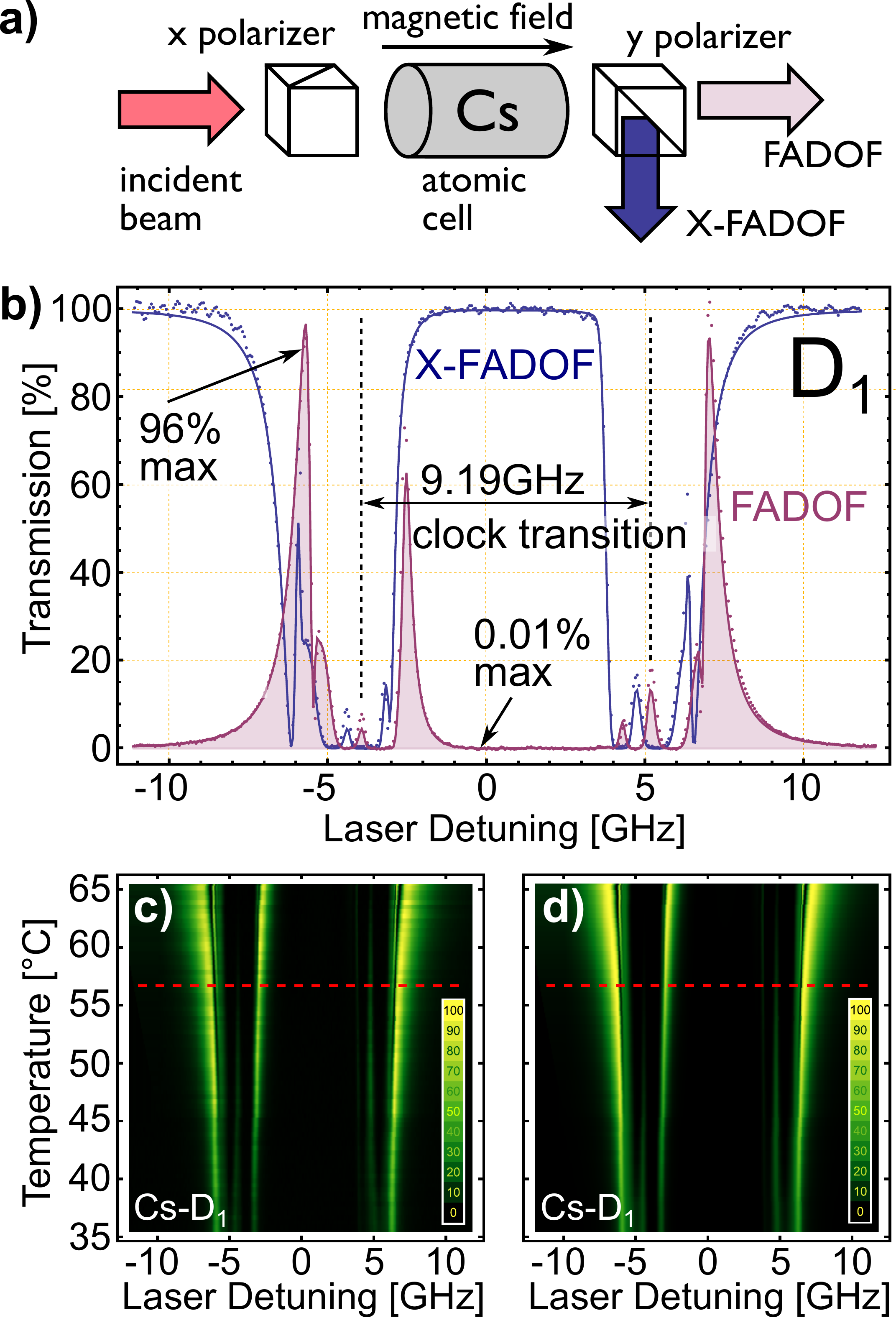}
  \caption{a) Experimental setup. A cesium vapor cell is placed between two crossed polarizers. A magnetic field is applied via a solenoid or permanent magnets. b) Transmission and absorption spectra of the Cs D$_1$-line plotted as function of the laser detuning. Transmission (FADOF, violet) and the second output port (X-FADOF, blue). When both spectra are added, the Doppler spectrum is obtained. Measurement data (dotted), fit (solid line). Center wavelength $\lambda_{\mathrm{vac}}$=894.593~nm c) Measured transmission as a function of temperature and laser detuning, red dashed line indicates data shown in b. d) Fit of the measured transmission data shown in d.}
  \label{fig:setup_comparison}
\end{figure}

The experimental configuration is depicted in Fig.~\ref{fig:setup_comparison}. A few $\mu$W of a narrow-band titanium-sapphire laser (TiSa, 899-21, Coherent, CA) are delivered by an optical single mode-fiber to the experiment. The beam-waist is $2 \cdot w_0$=4.5~mm. The frequency detuning is monitored by a Fabry-P\'{e}rot cavity and the incident laser power into the experiment is monitored by an amplified photo diode (not shown). The polarization state of the light is then fixed with a Glan-Laser polarizer (Thorlabs GT10-B) and passes through an evacuated anti-reflection coated cesium vapor cell (Triad Technologies, CO), with an optical path length of 100~mm. The cell was externally heated with copper blocks at the optical windows by a feed-back loop controlled heater. The temperature stability is estimated to be better than 0.5\degree C.

The FADOF transmission was analyzed by photo-diodes behind a Glan-Taylor calcite prism. Both output ports are monitored. In sum they represent the ordinary Doppler spectrum. For convenient data acquisition, all signals were recorded with an oscilloscope, triggered by the lasers control box. A full data set was acquired with temperatures ranging from 35\degree C to 70\degree C. The magnetic field of the solenoid was changed from 0 to 40~mT. The upper current limit was defined by the maximum voltage (32~V) of the power supply and the resistance of the solenoid, which amounts to 23.6~$\Omega$ for approx. 3000 windings.

To reduce the experimental complexity of the filter, and to avoid additional heating from the solenoid, we introduced permanent ferrite magnets instead of the solenoid. Eight axial magnetized ring-magnets (100$\times$60$\times$20~mm$^3$) result in an equal homogeneous longitudinal magnetic field of 37.5~mT. The ferrite based magnets retain their magnetization until 250\degree C. A comparable configuration was reported earlier, but in a high-field Hallbach configuration~\cite{rudolf_ol_2012}.
For a theoretical description, the FADOF is modeled by calculating the electric susceptibilities of the cesium vapor~\cite{yeh_ao_1982}. This is performed by calculating the Hamiltonian $H=H_0+H_{\mathrm{HFS}}+H_{\mathrm{Zeeman}}$ and results in the complex susceptibilities $\chi_{\pm}=\chi'_{\pm}+i \chi''_{\pm}$. Together with the length of the cell, $L$, and the temperature, a Voigt line profile is obtained per transition. The combination of optical rotation by $\pi/2$ and simultaneous weak Doppler absorption leads to a preferred optical configuration. The transmission is calculated as 

\begin{align}
T=&1/4 \biggl[ \exp\left(-\frac{\omega}{c}\chi''_{+}L\right)+\exp\left(-\frac{\omega}{c}\chi''_{-}L\right)\\
&-2\exp\left(-\frac{\omega}{c}\frac{\chi''_{+}+\chi''_{-}}{2}L\right)\cdot \cos\left(\frac{\omega}{c}\frac{\chi'_{+}-\chi'_{-}}{2}L\right)\biggr] \,. \nonumber
\end{align}
 
\noindent
We assume the Doppler broadening and the atomic density based on the same single temperature. A more detailed study of the theoretical estimated spectra is given in e.g.~\cite{harrell_josab_2009}.

For the theoretical model a self-developed program~\cite{gerhardt_w_2015} and additionally the package ``ElecSus''~\cite{zentile_cpc_2015} was used. For fitting the data ``ElecSus'' was preferred for providing a fast, Python-based, fitting routine. The frequency axis was linearized by fitting the Fabry-P\'{e}rot etalon transmission with an Airy-function. 

Fig.~\ref{fig:setup_comparison}b,c and d show the acquired FADOF-spectra along with the theoretical prediction.  For each spectrum a small background from ambient light was subtracted. Each spectrum was independently fitted. First, both temperature and B-field were set as free fit parameters. For the permanent magnet, a constant magnetic field 37.5$\pm$0.5~mT was found for each temperature. This value is in good agreement with the value obtained by a Hall sensor, and was set to a fixed value. The temperature axis plotted in Fig.~\ref{fig:setup_comparison}c and d, was obtained as a free fit parameter. The fit of each polarization axis provided us the same temperature. However, a small deviation (0.4\degree C) of the actual set point of the controller and the fit result was found.

The intention of the study was to find ways to optimally combine the properties of atomic vapors with the single photon emission of a QD. A few studies have been showing such an optical interaction. Single photons, originating from a single molecule have been filtered with a FADOF already~\cite{siyushev_n_2014}. With the presented data, this is the first experiment which should be applied to single photon emission from a QD. In the following we extend on the general filtering idea with two possible experiments:
 
At high excitation powers, the spectral emission of a single photon emitter is described by the dressed-state approach. The Mollow-triplet~\cite{mollow_pr_1969} introduces two side-bands, which are split by the Rabi frequency $\Omega$. This quantum optical feature has been experimentally realized in the solid state in the past~\cite{muller_prl_2007,wrigge_np_2008}. The emission on the side-bands is anti-correlated~\cite{aspect_prl_1980,ulhaq_np_2012}, since each part of the split ground-state can only be reached by an emission of the intermediate frequency. Only then, the other side-band is emitted. We find the FADOF as a convenient way to suppress the resonant scattering and to filter solely for the introduced side-bands. 

\begin{figure}[ht]
  \includegraphics[width=0.9\columnwidth]{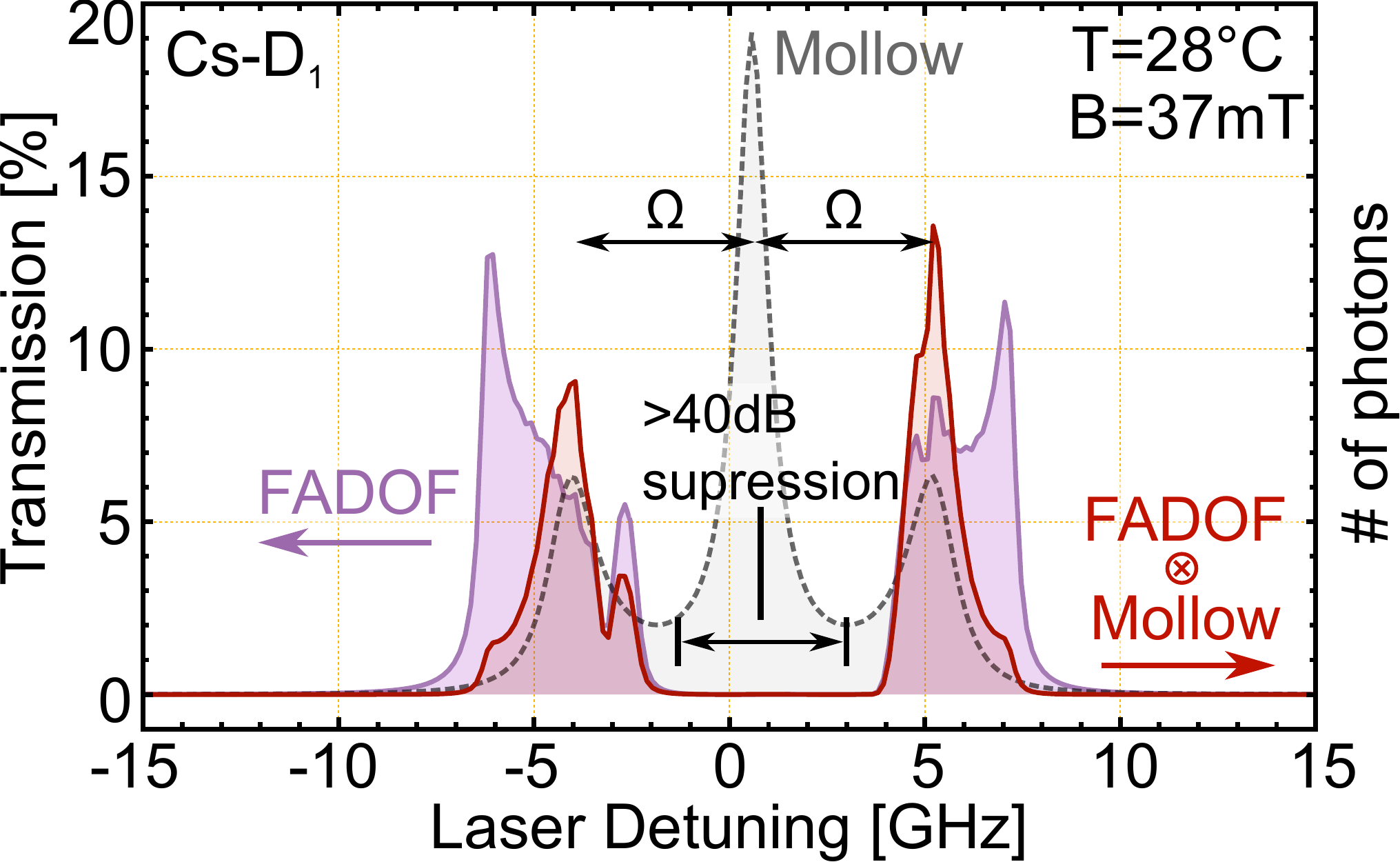}
  \caption{Spectrum of the Mollow triplet filtered by a Cs-D$_1$ FADOF as a function of the laser detuning. Mollow emission (gray, dashed), FADOF transmission (violet), convolution (red). The alternating emission from the sidebands coincides with the Cs clock-transition, whereas the middle-peak is efficiently suppressed.}
  \label{fig:fadofmollow}
\end{figure}

As one interesting side effect, the Mollow triplet might exhibit its side bands coinciding with the Cs-clock transition ($\approx$ 9.2~GHz). When the FADOF filter is used, the split photonic emission will be alternating passed, whereas the Rayleigh middle peak is suppressed. A general measure for the filter performance, can be given by a figure of merit (FOM), which is the ratio of maximum transmission and the background passing the filter~\cite{kiefer_s_2014} 

\begin{equation}
\textrm{FOM}= \frac{T_{\textrm{max}}}{\textrm{ENBW}} \qquad \textrm{ENBW} = \frac{\int T(\nu)d\nu}{ T(\nu_{\mathrm{S}})}
\end{equation}

\noindent
where $T_{\textrm{max}}$ is the maximum transmission. The equivalent noise band-width (ENBW) describes the equivalent noise bandwidth as a function of the transmission $T$ at a signal frequency $\nu_{\mathrm{S}}$. ENBW is the inverse of the signal to noise ratio for white noise. By altering certain parameters, e.g.\ temperature and B-field, the FOM can be maximized for the best filter performance~\cite{kiefer_s_2014}. However, by the given signal from a QD, the transmission of the sidebands through the filter needs to be optimized, whereas the resonant scattering (middle-peak) needs to be suppressed. Hence a convolution of transmission and Mollow-triplet sidebands is mandatory. This is presented in Fig.~\ref{fig:fadofmollow}. By varying both parameters, temperature (20-200\degree C) and B-field (0-50~mT), we find $T_{\mathrm{max}}$ between 20 and 40\degree C and above 20~mT. It is important to stay below 40\degree C, otherwise the middle-peak of the Mollow-triplet will pass the filter. Transmission of the entire Mollow-triplet is increased until a magnetic field of 20~mT. At higher fields, the transmission does not rise significantly. By the given magnetic field of 37.5~mT (permanent magnets) the best temperature for a maximized transmission of the sidebands is found to be 28\degree C. Simultaneously, the suppression of the center is better than 70~dB over a range of larger than 4.1~GHz. Assuming 10$^5$ counts per second (cps.) emitted by a QD, distributed over all three transitions, around 2,500 cps.\ will pass the filter in total. Therefore the filter has an transmission efficiency of 1.2~\% per sideband. This value is similar to the efficiency achieved with a Michelson interferometer~\cite{ulhaq_np_2012}. The Michelson interferometer for filtering the QD emission, however, is thermally and mechanically very susceptible. The Cs-FADOF is not only a more robust alternative, but can also be used to lock the emission of the QD.

\begin{figure}[b]
  \includegraphics[width=0.9\columnwidth]{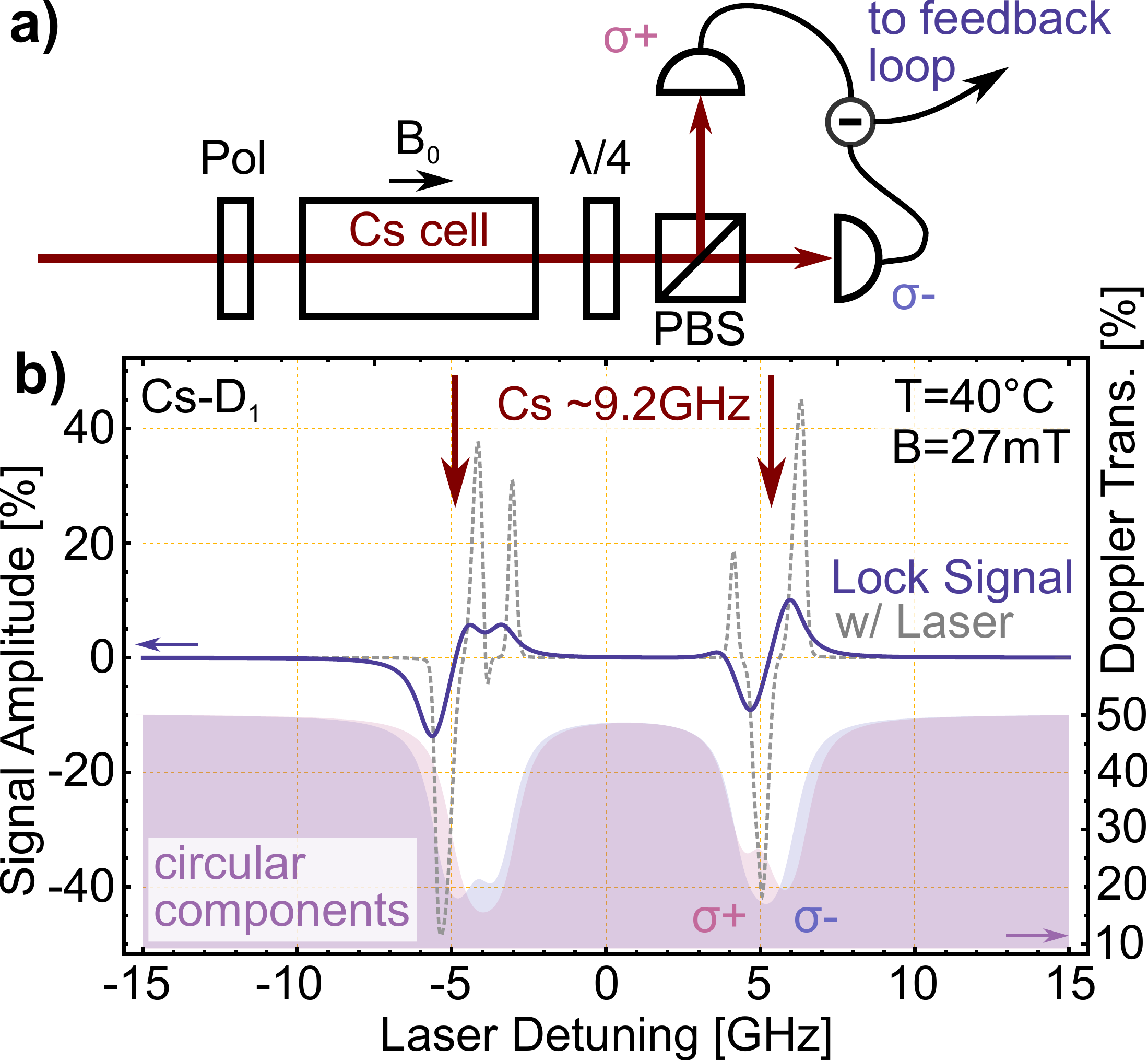}
  \caption{a) Experimental arrangement of the proposed locking scheme. The QD emission passes the Cs cell and both circular components are analyzed. b) Calculated lock signal with a laser (dashed) and with a QD (blue, linewidth of 500~MHz). Bottom: both detected signals on the photo detectors, top line represents 50\% transmission, bottom line is 10\% transmission. The locking bandwidth depends on the signal strength and the slope of the lock signal.}
  \label{fig:locking}
\end{figure}

Locking a QD, which only shows a single transition (unlike the Mollow triplet) to an atomic transition is important for further experiments, and for future applications in quantum information processing. Locking ensures, that the single photon emitter stays spectrally tuned to a follow up experiment with atoms. Tuning of the QD emission can be achieved for example by magnetic fields or the Stark effect. A first locking scheme has been presented by Akopian~et~al.~\cite{akopian_apa_2013}, but utilized a low locking bandwidth on the order of seconds. However, a single emitter can be conveniently locked by a dispersive lock signal, and by the comparison of only two signals. A slight modification of the presented FADOF scheme is required: As the incident linear-polarized light is a superposition of both circular components $\sigma_+$ and $\sigma_-$, these can be independently analyzed, when a quarter waveplate is introduced (see Fig.~\ref{fig:locking}a). This is comparable to the dichroic atomic vapor laser lock~\cite{petelski_tepjd-amoapp_2003}. Both output ports are equipped with single photon detectors. The signals are subtracted and the difference of zero defines the lock-point. If the signal deviates from zero, the single photon emitter has to be spectrally shifted.

Fig.~\ref{fig:locking}b shows the derived lock signal. Here, a spectral line width of 500~MHz for the single photon emitter is assumed. With a single photon emission of 10$^5$ cps., this implies that both detectors see far off-band the half of this value. At the atomic lock point it self, the rate is reduced to approx.\ 20\% by Doppler absorption. This is 10,000~cps. in each arm. The signal to noise ratio of the difference signal is therefore SNR(1sec)$=\sqrt{10,000}/\sqrt{2}$. Due to the fact that we have two detectors and subtract the signal, we have to introduce the factor of $\sqrt{2}$. For the stability, the slope of the lock signal becomes important. With the values as shown in Fig.~\ref{fig:locking}b we estimate this as a contrast change of 400 photons per MHz at the lock point. The lock precision, defined as $\delta \nu_{\mathrm{min}}(t)=\frac{\sigma(t)}{\delta D/\delta \nu}$, with a slow drifting emitter will be on the order of a MHz/$\sqrt{\mathrm{Hz}}$.


In summary, we have presented a study on the transmission of a Faraday filter on the cesium D$_1$-line. We find an excellent match between the theoretical prediction and the experimental data for the entire parameter space. This completes the study on alkali Faraday filters (except Li) and allows to apply the here presented results to single photon experiments with single solid state emitters. Since such experiments require sophisticated instrumentation, a solution based on a permanent magnet was introduced. 

For experiments with QDs, not only a simple superimposed spectrum is of interest. The suppression of the Rayleigh central line of the Mollow-triplet and locking the emitter to an atomic line extend the study and will allow for further experiments. In addition, experiments with slow light are of interest. One option would be to analyze the temporal shift between both output ports of such a filter. This compares to the early studies of slow light~\cite{grischkowsky_pra_1973}.

When preparing this manuscript, we found another study presenting results on the cesium D$_1$-line~\cite{zentile_a_2015}.

\textbf{Acknowledgement:} We acknowledge the funding from the MPG via a Max Planck fellowship (J.W.), the SFB project CO.CO.MAT/TR21, the Bundesministerium f\"ur Bildung und Forschung (BMBF), the project Q.COM, and SQUTEC.


%

\end{document}